\documentclass[11pt,a4paper]{article}
\usepackage{jcappub}

\usepackage{graphicx}
\usepackage{amsmath}
\usepackage{epsfig,multicol,bbm}

\newcommand{\be}{\begin{equation}}
\newcommand{\ee}{\end{equation}}
\newcommand{\bea}{\begin{eqnarray}}
\newcommand{\eea}{\end{eqnarray}}
\newcommand{\mbb}{\mathbb}

\newcommand{\mc}{\mathcal}

\newcommand{\beqa}{\begin{eqnarray}}
\newcommand{\eeqa}{\end{eqnarray}}

\newcommand{\vo}{\mathcal{V}}

\newcommand\fverb{\setbox\fverbbox=\hbox\bgroup\verb}
\newcommand\fverbdo{\egroup\medskip\noindent
			\fbox{\unhbox\fverbbox}\ }
\newcommand\fverbit{\egroup\item[\fbox{\unhbox\fverbbox}]}
\newbox\fverbbox

\title{Poly-instanton Inflation}

\author[a]{Michele Cicoli,}
\author[b]{Francisco G. Pedro,}
\author[c]{Gianmassimo Tasinato}

\affiliation[a]{Abdus Salam ICTP, Strada Costiera 11, Trieste 34014, Italy}
\emailAdd{mcicoli@ictp.it}
\affiliation[b]{Rudolf Peierls Centre for Theoretical Physics, University of Oxford,\\ 1 Keble Road, Oxford, OX1 3NP, United Kingdom}
\emailAdd{f.pedro1@physics.ox.ac.uk}
\affiliation[c]{Institute of Cosmology \& Gravitation, University of Portsmouth, \\
Dennis Sciama Building, Portsmouth, PO1 3FX, United Kingdom}
\emailAdd{gianmassimo.tasinato@port.ac.uk}

\abstract{We propose a new inflationary scenario
in type IIB Calabi-Yau compactifications,
where the inflaton is a K\"ahler modulus parameterising the volume of an internal four-cycle.
The inflaton potential is generated via poly-instanton corrections to the superpotential which
give rise to a naturally flat direction due to their double exponential suppression.
Given that the volume mode is kept stable during inflation,
all the inflaton-dependent higher dimensional operators are suppressed.
Moreover, string loop effects can be shown
to be negligible throughout all the inflationary
dynamics for natural values of the underlying parameters. 
The model is characterised by a reheating temperature
of the order $T_{\rm rh}\simeq 10^6$ GeV which requires $N_e \simeq 54$ e-foldings of inflation.
All the inflationary observables are compatible with current observations
since the spectral index is $n_s \simeq 0.96$,
while the tensor-to-scalar ratio is $r\simeq 10^{-5}$.
The volume of the Calabi-Yau is of order $10^3$ in string units,
corresponding to an inflationary scale around $10^{15}$ GeV.}

\begin{document}
\maketitle

\section{Introduction}

It is now 30 years since cosmic inflation was proposed to solve outstanding problems
of the standard cosmological model by postulating a period
of exponential expansion in the early Universe \cite{Guth:1980zm,Linde:1981mu,Albrecht:1982wi}.

Since the proposal of the original ideas, a plethora of inflationary models was put forward,
among which we find the class of slow-roll inflationary models.
At their heart is a slowly evolving field in a nearly flat potential that generates
a quasi-de Sitter phase in the early history of the Universe. Due to its simplicity,
the slow roll mechanism is among the preferred ways to generate an inflationary epoch.
While as a class of models, slow roll provides a successful and elegant realisation of inflation,
pinning down the details of the mechanism is a challenging question.
In particular, identifying the inflaton field, and its connections
with a fundamental theory of particle physics,
are still questions open to debate. The best that can be done is to propose
and analyse models that are compatible with current observational and experimental bounds,
and compute their signatures so that they can be tested by forthcoming observations.

Moreover, the very shallow inflaton potential yields a scalar mass lighter than the
Hubble scale $H$, $m_{\phi} \ll H$. However, as in the Higgs case,
it is notoriously hard to keep scalars light by preventing them from
getting large contributions when integrating out heavy ultra-violet physics.
This is the famous `$\eta$-problem' \cite{etaproblem} whose solution is
crucial in order to trust any inflationary scenario.
Due to this ultra-violet sensitivity of inflation,
it is possible to find a robust solution to the $\eta$-problem
only by embedding models in an ultra-violet complete theory.

For these reasons, a promising avenue to embed inflation into a model of particle physics
is string theory (for reviews see \cite{Reviews,OurReview}).
A key feature of string theory is the need for extra spacetime dimensions whose geometry is parameterised by scalar moduli fields.
It is the vacuum structure of the moduli potential that determines the masses and couplings
of the low-energy effective field theory. Over the past decade, significant progress has been made
towards the understanding of the moduli potential \cite{kklt,LVS},
allowing for a promising contact between string theory and particle phenomenology.
The progress in moduli stabilisation also opened up the possibility of realising inflation in the moduli sector.
String inflationary models based on single-field slow-roll can be broadly classified,
based on the origin of the inflaton field \cite{Openclosedinflatons}, into open string models \cite{DDinflation,BBbarInfl,DBI}
and closed string models \cite{Racetrack,KMI,FI,Nflation,SW,nemanja,VolInfl,OtherClosedStringInfl}.

For closed string inflation there is a direct connection between the physics that stabilises the extra dimensions and inflation.
Of particular interest for string phenomenology and inflationary applications
is the LARGE Volume Scenario (LVS) of type IIB string theory \cite{LVS}
that through a combination of perturbative and non-perturbative effects allows for solutions with exponentially large volumes.
In the context of the LVS, inflation can be driven by K\"ahler moduli rolling towards their minima.

As recently reviewed in \cite{OurReview}, these models provide an interesting
solution to the $\eta$-problem which does not rely on an axionic shift symmetry \cite{shiftInflation}.
The two reasons why these models can evade the $\eta$-problem can be summarised in the following way:
\begin{enumerate}
\item The characteristic no-scale structure of the K\"ahler potential of type IIB supergravity is broken
at leading order by $\alpha'$ effects which develop a potential \textit{only} for the
overall volume mode. Thus all the other $(h^{1,1}-1)$ directions in the K\"ahler moduli
space orthogonal to the volume are flat at leading order, and so constitute natural inflaton candidates.
The `extended no-scale structure' guarantees that string loop effects generate only a subleading potential
lifting some of the remaining flat directions \cite{Cicoli:2007xp}. Hence any possible direction orthogonal to the volume is a good inflaton candidate.

\item The tree-level K\"ahler potential $K$ itself depends \textit{only} on the overall volume.
 Therefore, if the inflaton is a combination of the K\"ahler moduli orthogonal to the volume mode (or, equivalently,
 if inflation takes place with the volume kept stable),
 no higher order inflaton-dependent operator gets generated by expanding the prefactor $e^K$
 of the F-terms scalar potential. Given that $g_s$ corrections to $K$ generically induce a dependence
 on all the K\"ahler moduli, inflaton-dependent higher order operators get indeed generated by expanding $e^K$
 once we consider the $g_s$-corrected K\"ahler potential. However, again due to the `extended no-scale structure',
 these operators are suppressed with respect to the leading inflationary dynamics.
\end{enumerate}

Two possible effects to develop a potential for the inflaton field at subleading order
have already been proposed. In the first case, K\"ahler moduli inflation \cite{KMI}, the inflaton is a blow-up mode
and its potential is generated by ordinary non-perturbative corrections to the superpotential,
while in the second case, fibre inflation \cite{FI}, the inflaton is a K3 divisor which develops
a potential via perturbative string loop effects \footnote{The authors of \cite{VolInfl}
tried to obtain inflation driven by the volume mode in order to solve the tension
between inflation and TeV-scale SUSY \cite{Kallosh:2004yh}. However, for the reasons explained above, this model
suffers from the $\eta$-problem, and so its realisation needs severe fine-tuning.}.
In fibre inflation a long enough period of inflation can be achieved rather naturally.
On the other hand, in K\"ahler moduli inflation, a sufficiently long period of inflation can be driven by the non-perturbative potential only
by fine-tuning the coefficients of the loop corrections, as pointed out in \cite{FI}. Notice that these $g_s$ corrections would spoil inflation in the region rather close to the minimum where
the non-perturbative potential would give rise to slow-roll. On the other hand,  in regions further away
from the minimum, due to the `extended no-scale structure', these  perturbative loop corrections
would actually drive inflation in a way very similar to \cite{FI}.

In this paper we shall provide the first model where the whole inflationary dynamics
is driven entirely by non-perturbative effects since perturbative loop corrections can be shown to be
negligible for natural values of the underlying parameters.
The new key-ingredient is the use of poly-instanton corrections
to the superpotential which are instanton
corrections to the action of another instanton derived in type I compactifications \cite{Blumenhagen:2008ji}.
The type IIB origin of these poly-instanton corrections has not been worked out in detail yet.
However, as argued in \cite{ADDstrings}, the poly-instantons might get generated either for a fluxed instanton on a K3 four-cycle
or for an instanton on a $T^4$ divisor with deformations fixed by the background fluxes.
We shall therefore focus on Calabi-Yau three-folds which admit a K3 or $T^4$ fibration over a $\mbb{P}^1$ base
similarly to the toric constructions of \cite{Cicoli:2011it}.

Here the inflaton is the volume of the fibre which supports the poly-instanton effects.
The different topological origin of the inflaton between our model and the one developed in \cite{KMI}
results in a different canonically normalised inflaton field.
This leads to a potential which in our case gives rise to slow-roll in a region much closer to the
minimum than in the model of \cite{KMI} even if in both cases the inflationary potential is
generated by non-perturbative effects. This is the reason why in our model string loop
corrections are less dangerous and can be shown to be negligible throughout all the inflationary
dynamics for rather natural values of the underlying parameters.

The predictions of the model are rather independent on the choice of the microscopic parameters,
as these influence mainly the scale of inflation but have little effect on the shape of the potential.
Our model is characterised by a reheating temperature of the order $T_{\rm rh}\simeq 10^6$ GeV
which requires $N_e\simeq 54$ e-foldings of inflation. The requirement of
generating the correct amount of density perturbations fixes the Calabi-Yau volume
of the order $10^3$ in string units. This, in turn, leads to a high inflationary scale, $M_{\rm inf}\simeq 10^{15}$ GeV,
corresponding to a small tensor-to-scalar ratio, $r\simeq 10^{-5}$, and a sub-Planckian
motion of the inflaton field $\Delta\phi\simeq 0.5 M_p$. We also point out that the spectral index, $n_s\simeq 0.96$,
is within the observationally allowed window.

This paper is organised as follows:
we start in section \ref{LVSReview} by reviewing the LARGE Volume Scenario of type IIB string theory
with particular emphasis on fibred constructions. In section \ref{Poly}
we first show how the poly-instanton corrections to the superpotential
can generate a potential for the fibre modulus, and then
we study its inflationary applications both analytically and numerically.
In the same section we also present a detailed discussion of the $\eta$-problem
stressing the similarities and the differences of our model
with the previous ones developed in \cite{KMI} and \cite{FI}, where the inflaton
was still a K\"ahler modulus.
We finally conclude in section \ref{Concl}.

\section{Basics of the LARGE Volume Scenario}
\label{LVSReview}

In this section we review the basic properties of the LVS, to be used in what follows
for cosmological purposes.

\subsection{4D effective action}

The low-energy 4D $\mc{N}=1$ supergravity, obtained by compactifying type IIB string theory
on a orientifold of a Calabi-Yau three-fold $\mc{M}$, is characterised by a plenitude of scalar fields
that parameterise the geometry of $\mc{M}$. These include the axio-dilaton $S$, the complex structure moduli $U_a$,
$a=1,...,h^{1,2}(\mc{M})$ and the K\"ahler moduli $T_i$, $i=1,...,h^{1,1}(\mc{M})$.
Due to the no-scale structure of the tree-level K\"ahler potential for the K\"ahler moduli \cite{Grimm:2004uq}:
\be
K_0= -2\ln\vo\,,
\label{eq:KnoScale}
\ee
and the fact that the semiclassical superpotential $W_0$,
generated by turning on non-zero background fluxes \cite{Gukov:1999ya, Giddings:2001yu},
does not depend on the $T$-moduli, only the $S$ and $U$-moduli are fixed supersymmetrically at tree-level
by imposing vanishing F-term conditions.
In fact, for a constant superpotential $W_0$, the F-term scalar potential for the K\"ahler moduli:
\be
	V=e^K (K^{i \bar{j}} D_i W D_{\bar{j}}\bar{W} -3 |W|^2)\,,
	\label{eq:V}
\ee
identically vanishes due to the no-scale structure relation following from (\ref{eq:KnoScale}):
\be
K_0^{i \bar{j}} \partial_i K_0 \partial_{\bar{j}}K_0=3\,.
\label{noscale}
\ee
The dependence of the Calabi-Yau volume $\vo$ on the K\"ahler moduli can be
obtained by recalling its definition in Einstein frame:
\be
\vo= \frac{1}{6} k_{ijk} t^i t^j t^k,
	\label{eq:vol}
\ee
where $k_{ijk}$ are triple intersection numbers and $t^i$ two-cycle volumes.
The fields appearing in the 4D effective theory are the complexified K\"ahler moduli $T_i\equiv \tau_i + i b_i$
where $b_i$ are axionic components while $\tau_i$ are four-cycle volumes, given by:
\begin{equation}
	\tau_i=\frac{\partial \mathcal{V}}{\partial t^i}=\frac{1}{2}k_{ijk}t^j t^k.
	\label{eq:4cycle2cycle}
\end{equation}
Inverting (\ref{eq:4cycle2cycle}), and then inserting this result back in (\ref{eq:vol}),
one can work out the dependence of the volume on the $T$-moduli.
As we shall discuss later on, the no-scale structure of the system (\ref{noscale})
has important positive consequences for building models of inflation.

In order to have a phenomenologically viable model it is essential to stabilise all the moduli.
Since the leading order terms in the K\"ahler potential and superpotential do not generate a minimum for the K\"ahler moduli,
one must go one step further and see how the inclusion of subleading terms in $K$ and $W$ changes the scenario.
These higher order corrections can be classified into perturbative and non-perturbative.

Since $W$ is protected by supersymmetric non-renormalisation theorems it is immune from
  perturbative corrections: therefore these can only occur in the K\"ahler potential.
The no-scale structure can be broken both by $\alpha'$ and $g_s$ effects.
The leading order $\mathcal{O}(\alpha'^3)$ correction to the K\"ahler potential has been computed
in \cite{BBHL} and takes the form:
\begin{equation}
		K=-2 \ln \left(\mathcal{V}+\frac{\xi}{2 g_s^{3/2}} \right)\,,
	\label{eq:KBBHL}
\end{equation}
where $\xi=-\frac{\zeta(3)\chi(\mathcal{M})}{2 (2 \pi)^3}$ with $\chi(\mathcal{M})=2\left(h^{1,1}-h^{2,1}\right)$.
String loop effects also break the no-scale structure by adding new contributions to $K$ \cite{08051029,Cicoli:2007xp}.

The inclusion of non-perturbative corrections to the superpotential
provides an alternative way of breaking the tree-level no-scale structure \cite{kklt}.
This is achieved by considering contributions to $W$ coming from gaugino condensation on D7-branes
or Euclidean D3-brane instantons:
\be
	W= W_0 + \sum_i A_i e^{- a_i T_i},
	\label{eq:W}
\ee
where $A_i$ and $a_i$ are independent from the K\"ahler moduli, although they
 depend on the microscopic details of the model.
 In what follows we will take them to be constants, with $a_i = 2 \pi/n_i$
 and $n_i$ natural numbers.

\subsection{Swiss-cheese compactifications}

In \cite{LVS} it was shown that the combination of $\mc{O}(\alpha'^3)$ corrections to $K$
 and non-perturbative effects in $W$ can generate an AdS minimum for the K\"ahler moduli
 at exponentially large volumes for geometries in which the volume can be written as:
\be
\mc{V}=\alpha\left(\tau_{1}^{3/2}- \sum_{i=2}^{h^{1,1}} \gamma_i\tau_{i}^{3/2}\right).
\label{eq:volCheese}
\ee
These are the so-called Swiss-cheese geometries in which one K\"ahler modulus
controls the overall volume while the remaining $(h^{1,1}-1)$ K\"ahler moduli
are the volumes of diagonal del Pezzo divisors. The scalar potential for these geometries behaves as:
\be
V=\sum_i \frac{8 (a_i A_i)^2 \sqrt{\tau_i} e^{-2a_i \tau_i}}{3 \mc{V} \alpha \gamma_i}
-\sum_i \frac{4 a_i A_i W_0 \tau_i e^{-a_i \tau_i}}{\mc{V}^2}+\frac{3 \xi W_0^2}{4 g_s^{3/2} \mc{V}^3}.
\ee
The minus sign in front of the second term is crucial for the existence of the minimum
and comes from the minimisation of the axionic potential.
Furthermore one must require $\xi>0$ to have a minimum for the potential
where the blow-up cycles are fixed at leading order at:
\be
\langle\tau_i\rangle \simeq \frac{1}{a_ i g_s} \left( \frac{\xi}{2 \alpha \gamma_i\sum_j \gamma_i/a_i^{3/2}}\right)^{2/3},
\label{eq:leadingVevBU}
\ee
while the volume is stabilised at:
\be
\langle\mc{V}\rangle\simeq\frac{3 \gamma_i \alpha W_0}{2 a_i A_i}\left(\frac{1-a_i\tau_i}{1/2-2a_i\tau_i}\right)
\sqrt{\tau_i} \,e^{a_i \tau_i}\,.
\ee
Notice that the combination of $\alpha'$ corrections to $K$ and non-perturbative effects in $W$
allows for the generation of a minimum for all the blow-up moduli plus the volume mode.
For Swiss-cheese type geometries this is sufficient to fix all the $T$-moduli \footnote{See \cite{Palti:2008mg}
 for an example in type IIA string theory.}.

\subsection{Fibred compactifications}

The existence of an AdS minimum at large volumes for
manifolds more general than Swiss-cheese geometries was studied in \cite{08051029}.
Of particular interest to us are Calabi-Yau manifolds that admit a K3 or $T^4$ fibration over a $\mathbb{P}^{1}$ base.
These manifolds are characterised by the fact that their volume is linear in the two-cycle giving the volume of the $\mbb{P}^1$ base.
Explicit examples of this kind of Calabi-Yau three-folds with additional del Pezzo divisors have been analysed in \cite{Cicoli:2011it}
using the powerful tools of toric geometry.
Here we shall just focus on the simplest of such manifolds whose volume takes the form:
\be
\vo=\lambda_1 t_1 t_2^2+\lambda_2 t_3^3\,,
\ee
where $t_1$ is the volume of the $\mbb{P}^1$ base, $\tau_1= \lambda_1 t_2^2$
is the size of the K3 or $T^4$ fibre, and $\tau_3=3 \lambda_2 t_3^2$
controls the volume of a blow-up mode (the other four-cycle volume
is given by $\tau_2=2\lambda_1 t_1 t_2$). Notice that the fibre is a
K3 surface if its Euler characteristic is $\chi=24$ whereas it is a $T^4$ divisor
if $\chi=0$ \cite{Schulz}. We shall not specify the value of $\chi$ in order to
be as generic as possible since our moduli stabilisation mechanism works in both cases.
We can then rewrite the volume in terms of the correct K\"ahler moduli as:
\be
\vo= t_1 \tau_1-\alpha\gamma \tau_3^{3/2} =\alpha(\sqrt{\tau_1}\tau_2 -\gamma\tau_3^{3/2})\,,
\label{eq:volFibre}
\ee
where $\alpha=1/(2\sqrt{\lambda_1})$ and $\gamma=\frac 23 \sqrt{\lambda_1/(3\lambda_2)}$.

Due to the similar structure of the two volume expressions, Eqs. (\ref{eq:volCheese}) and (\ref{eq:volFibre}),
an analysis of the scalar potential reveals that an AdS minimum at exponentially large volumes also exists for fibred geometries.
There is however one crucial difference: while in the simpler geometries initially studied, with volumes described by Eq. (\ref{eq:volCheese}),
all moduli were stabilised, for K3 or $T^4$-fibered Calabi-Yau manifolds
only the blow-up moduli and the volume $\mc{V}\sim \sqrt{\tau_{1}}\tau_{2}$ directions are fixed.
Therefore the leading order stabilisation dynamics leaves one flat direction
in moduli space which can be used to provide a stringy realisation of slow roll inflation.

Before proceeding to study the inflationary dynamics, we finally point out that
in this discussion we neglected many aspects related to the construction of a
globally consistent compactification with the local presence of chiral matter.
We shall however assume that the problems that one generically encounters when
trying to combine moduli stabilisation with D-brane model building can be cured
along the same lines described in \cite{CMV}.

\section{Poly-instanton inflation}
\label{Poly}

In this section we introduce poly-instanton corrections to the superpotential. These are the essential ingredient to lift the flat direction
left over after the study of the leading stabilisation dynamics. We then show that the resulting potential for the fibre modulus
can support slow-roll inflation. We provide both analytic arguments and numerical results for several representative points in moduli space.

\subsection{Fibre stabilisation}

The existence of poly-instanton corrections was first established within the context of type I compactification in \cite{Blumenhagen:2008ji}.
These are instantons which do not give rise to a single contribution to the superpotential but,
due to the presence of extra fermionic zero-modes, they correct the action of another instanton wrapping a different internal cycle.
The zero-mode constraints for generating poly-instanton corrections in the type IIB T-dual version has not been worked out in detail yet
and it is beyond the scope of this paper. We shall therefore take a phenomenological approach and
make the following two assumptions:
\begin{enumerate}
\item The field theory living
on a stack of D7-branes wrapping $\tau_3$ can be broken into two gauge groups which separately undergo gaugino condensation,
giving rise to a race-track superpotential of the form:
\be
W=W_0+A \,e^{-a T_3} -B \,e^{-b T_3}\,,
\label{eq:Wracetrack}
\ee
where $A$ and $B$ are threshold effects, while $a=2 \pi/n_a$ and $b=2 \pi/n_b$, with $n_a, n_b\in \mbb{N}$.

\item On top of these effects, a Euclidean D3-instanton wrapping the fibre $\tau_1$ yields non-perturbative corrections
to the gauge kinetic functions of the two condensing gauge groups, resulting in a poly-instanton corrected
superpotential which looks like:
\be
W=W_0+A \,e^{-a (T_3+C_1 e^{-2 \pi T_1})} -B \,e^{-b (T_3+C_2 e^{-2\pi T_1})}\,,
\label{eq:Wpoly}
\ee
where $C_1$ and $C_2$ are free constants.
\end{enumerate}
Notice that we are following \cite{ADDstrings}, where the same assumptions led to
stabilise the fibre four-cycle obtaining a very anisotropic compactification characterised by
the presence of two micron-sized extra dimensions and strings around the TeV scale.
The authors of \cite{ADDstrings} pointed out that our set-up is very likely to generate
poly-instanton corrections to the superpotential since the two extra fermionic zero-modes
of the type I computation are Wilson line modulini and these get mapped
to Wilson line and deformation modulini in the type IIB picture. Hence,
given that a K3 fibre admits deformations while a $T^4$ surface has Wilson lines,
it is not unreasonable to expect that these poly-instanton effects can indeed
get generated in our set-up either for a magnetised instanton on K3 or for an ED3 on $T^4$
with deformations fixed by the same background fluxes turned on to stabilise the
dilaton and the complex structure moduli \cite{ADDstrings}.

The scalar potential computed by using Eq. (\ref{eq:V}) can be separated as:
\be
V=V_{\mc{O}(\mc{V}^{-3})}+V_{\mc{O}(\mc{V}^{-3-p})}\,,
\ee
where the first piece scales with the volume as ${\mc V}^{-3}$, while the other
as ${\mc V}^{-3-p}$, with $p$ a positive parameter to be defined later.
After minimising with respect to the two axions $b_1$ and $b_3$, these two pieces are:
\be
\begin{split}
V_{\mc{O}(\mc{V}^{-3})}=&\frac{8 \sqrt{\tau_3}\left(A^2 a^2 e^{-2 a \tau_3}+B^2 b^2 e^{-2 b \tau_3} -2A B \, a b\,
 e^{- (a +b)\tau_3}\right)}{3 \mc{V}}\\
&+\frac{4 W_0 \tau_3 \left(A a e^{- a \tau_3}-B b e^{- b \tau_3}\right)}{\mc{V}^2}+\frac{3 \xi}{4 \,g_s^{3/2}\,\mc{V}^3}\,,
\end{split}
\label{eq:Vleading}
\ee
and:
\be
\begin{split}
V_{\mc{O}(\mc{V}^{-3-p})}=&-e^{-2 \pi \tau_1}\left\{ -\frac{16 C_1\sqrt{\tau_3}e^{-2 b\tau_3}}{3 \mc{V}}[Z^2 a-Z B (a-b)]\right.\\
&\left.+\frac{4 C_1 W_0 e^{-b\tau_3}}{\mc{V}^2}[2 \pi\,Z  \tau_1+Za\tau_3-B b(a-b)\tau_3]\right.\\
&\left.+n\left[\frac{-16 B b^2 \sqrt{\tau_3}e^{-2 b \tau_3}Z}{3\mc{V}}+\frac{4 W_0 B b e^{-b\tau_3}}{\mc{V}^2}(b \tau_3+c\tau_1)\right]\right\}\,.
\end{split}
\label{eq:Vsubleading0}
\ee
In Eq. (\ref{eq:Vleading}) we kept only the leading order terms in the volume expansion, in particular we consistently neglected terms proportional to $\mc{V}^{-n}$ for $n>3$ and terms proportional to $e^{-2\pi \tau_1}$.  In Eq. (\ref{eq:Vsubleading0}) terms proportional to $e^{-2\pi n\tau_1}$ for $n>1$ were left out as they are subdominant in the regime where the effective field theory is under control. This is equivalent to expanding the double exponentials of Eq. (\ref{eq:Wpoly}) as products of single exponentials.
Moreover, following \cite{ADDstrings},
we also introduced new parameters to simplify the expressions:
\be
Z\equiv B\, b-A\,a\,e^{-\left(a-b\right) \tau_3}\qquad \text{and}\qquad n\,\equiv\,C_2-C_1\,.
\label{eq:DefnZ}\,
\ee
The leading contribution, Eq. (\ref{eq:Vleading}), is the standard LVS potential obtained
from a racetrack superpotential. Notice that this term carries no information about the poly-instantons in $W$.
Minimisation with respect to the volume and the blow-up modulus provides a minimum at:
\be
\langle \tau_3 \rangle^{3/2}=\frac{3\xi}{32\alpha\gamma f_1(1-f_1) g_s^{3/2}}\,,
\qquad \langle \vo \rangle= f_2 \,e^{b \langle\tau_3\rangle}\,,
\label{eq:LVSMmin}
\ee
where we have defined:
\be
f_1\equiv \frac{\left(a-b\right) \tau_3 B b+ Z (1-a\,\tau_3  )}{4 \left(a-b\right) \tau_3 B b+Z(1-4 a\,\tau_3)}\,,
\qquad \text{and}\qquad f_2\equiv\frac{3 \alpha \gamma W_0 \sqrt{\langle\tau_3\rangle}}{ Z} f_1\,.
\ee
We stress that the result of Eq. (\ref{eq:LVSMmin}) involves no approximations and is therefore exact.
In particular the function $f_1$ generalises the definition of $f_{\rm corr}$ in \cite{ADDstrings}.
The equation that determines the VEV of $\tau_3$ can be solved iteratively,
giving a more accurate result than the leading order estimate obtained by setting $f_1=1/4$.
Notice that the volume is stabilised at exponentially large values, leading to a LVS.

As we have already mentioned, the leading order potential,  Eq. (\ref{eq:Vleading}),
depends on $\tau_1$ only through $\mc{V}$, therefore after stabilising $\tau_3$ and $\mc{V}$ at their minima,
Eq. (\ref{eq:LVSMmin}), there is one flat direction left in the $(\tau_1,\tau_2)$ plane.
This flat direction is then lifted by the subleading effects due to poly-instantons encoded in Eq. (\ref{eq:Vsubleading0}).
Notice that unlike $V_{\mc{O}(\mc{V}^{-3})}$ which depends only on $\tau_3$ and $\mc{V}$,
$V_{\mc{O}(\mc{V}^{-3-p})}$ depends on all the three directions in moduli space.
The approach of using subleading non-perturbative corrections to the superpotential as in \cite{ADDstrings}
provides an alternative for generating a minimum for the fibre modulus,
with respect to using string loop contributions to the K\"ahler potential
as originally explored in \cite{08051029}.

Setting the blow-up and volume moduli at their minima, Eq. (\ref{eq:LVSMmin}),
we can rewrite the subleading part of the
resulting scalar potential in Eq. (\ref{eq:Vsubleading0}) as:
\be
V_{\mc{O}(\mc{V}^{-3-p})}=-\frac{F_0}{\langle\vo\rangle^3} \left(2\pi\,\tau_1-p\, b\, \langle\tau_3\rangle\right)
e^{-2\pi\, \tau_1}\,,
\label{eq:Vsubleading}
\ee
where $F_0\equiv 4\,f_2\,W_0\,r_1$ and $p$ is defined in terms of the other parameters as:
\be
p\,\equiv\,-\frac{r_1}{r_3}\,,
\label{Defp}
\ee
with:
\bea
r_1&\equiv& C_1 Z+n \,b\, B\,,
\label{eq:Defnr1} \\
r_3&\equiv& (1-4 f_1)\left[ \frac{a}{b}\,r_1- (a-b)\,B\,(C_1+n)\right]\,.
 \label{eq:Defnr3}
\eea
The minimum of $V_{\mc{O}(\mc{V}^{-3-p})}$ for fixed $\tau_3$ and $\mc{V}$ can be shown to lie at:
\be
2\pi\,\langle\tau_1\rangle\,=\,p\, b\, \langle\tau_3\rangle+1.
\label{minimum}
\ee
By tuning the parameters in $K$ and $W$, it is possible to obtain $p$ positive and of order unity so that
the minimum (\ref{minimum}) is within the regime of validity of the effective field theory.
As we have already said, ref. \cite{ADDstrings} obtained
a very anisotropic configuration with TeV-scale strings corresponding to volumes of the order $\vo \sim 10^{30}$.
Here we shall instead be interested in a higher fundamental gravity scale which is suitable for cosmological purposes,
and so we shall choose our underlying parameters in order to obtain smaller volumes of the order $\vo \sim 10^3$.

\subsection{Inflationary potential}

Let us now study the inflationary dynamics working in the single field approximation.
We shall set the blow-up mode $\tau_3$ and the volume $\vo$ at their
$\tau_1$-independent VEVs: $\tau_3=\langle\tau_3\rangle$ and $\vo=\langle\vo\rangle$,
and displace $\tau_1$ far from its minimum. We shall then investigate if
the fibre modulus can drive a long enough period of inflation while
rolling down its potential. Notice that the approximation of stable $\tau_3$ and $\vo$ during inflation
is justified by the mass hierarchy found in \cite{ADDstrings}:
\be
m_{\tau_1}\sim \frac{M_p}{\mc{V}^{(3+p)/2}}, \qquad
m_{\vo}\sim \frac{M_p}{\vo^{3/2}}, \qquad m_{\tau_3}\sim \frac{M_p}{\vo}\,,
\ee
when all the moduli sit at their minima.
In what follows we shall set $p\simeq \mc{O}(1)$, implying that
for large volume $m_{\tau_3}\gg m_{\vo}\gg m_{\tau_1}$.
Moreover, as we shall see later on, the Hubble parameter
during inflation scales with the volume
as $H \sim m_{\tau_1}$. Hence both $\tau_3$ and $\vo$ are heavy during inflation since
$m_{\tau_3}\gg m_{\vo}\gg H$, justifying our single field approximation.
We point out that the actual mass of $\tau_1$ far from its minimum is much
smaller than $m_{\tau_1}$ due to the rapid exponential suppression of its potential.
Hence if we displace this field far from its minimum,
it is naturally lighter than $H$, and can therefore drive inflation.

The effective potential for the inflaton field $\tau_1$ can be written as:
\be
V_{\rm inf}=V_{\rm up}+V_{\rm fib}\,,
\ee
where $V_{\rm fib}$ is the scalar potential (\ref{eq:Vsubleading}) generated by poly-instanton effects,
while $V_{\rm up}$ is the uplift term. Regardless of the nature of the uplifting,
we shall assume that it gives rise to a $\tau_1$-independent constant once the volume is fixed.
This constant is obtained by requiring $V_{\rm up}=-V_{\rm fib}(\langle \tau_1\rangle)$
since we are tuning the uplifting in order to ensure the vanishing of the scalar potential
at leading $\vo^{-3}$ order. Notice that $V_{\rm up}$ will slightly modify the
position of $\langle \vo \rangle$ but we shall neglect this small effect since
it only affects the overall inflationary scale.

Let us then rewrite the full inflationary potential as:
\be
V_{\rm inf}=\frac{F_{\rm poly}}{\langle\vo\rangle^{3+p}}
\left[1-(1+2\pi \hat \tau_1)\,e^{-2\pi \hat \tau_1}\right]\,,\quad\text{with}\quad F_{\rm poly}=F_0 f_2^p\,e^{-1}\,,
\label{eq:Vinf}
\ee
where we defined $\hat \tau_1\equiv \tau_1-\langle\tau_1\rangle$ as
the shift of the inflaton from its minimum.
Notice that the inflationary potential scales with the volume as $1/{\cal V}^{3+p}$,
and so we can get different values for the scale of inflation by varying the parameter $p$.
This will be important in what follows.

Let us now canonically normalise the inflaton field by recalling that the kinetic Lagrangian
is given by:
\be
\mc{L}_{\rm kin}=K_{i\bar{j}}\partial_{\mu} T_i \partial^{\mu} \bar{T}_j
=\frac{1}{4}\frac{\partial ^2K }{\partial \tau_i \partial \tau_j}
(\partial_{\mu} \tau_i \partial^{\mu} \tau_j+\partial_{\mu} b_i \partial^{\mu} b_j)\,,
\ee
where in the second equality we rewrote the complexified K\"ahler moduli in terms of its real and imaginary components ($T_i=\tau_i+i b_i$) and used the fact that the K\"ahler potential is independent of the axionic fields.
The K\"ahler metric turns out to be:
\begin{equation}
K^0_{i \bar{j}}=\left(
\begin{array}{ccc}
\frac{1}{4 \tau _1^2} & \frac{\tau _3^{3/2}}{4 \tau _1^{3/2} \tau _2^2} & -\frac{3 \sqrt{\tau _3}}{8 \tau _1^{3/2} \tau _2} \\
 \frac{\tau _3^{3/2}}{4 \tau _1^{3/2} \tau _2^2} & \frac{1}{2 \tau _2^2} & -\frac{3 \sqrt{\tau _3}}{4 \sqrt{\tau _1} \tau _2^2} \\
-\frac{3 \sqrt{\tau _3}}{8 \tau _1^{3/2} \tau _2} & -\frac{3 \sqrt{\tau _3}}{4 \sqrt{\tau _1} \tau _2^2} &\frac{3}{8 \sqrt{\tau _1} \tau _2 \sqrt{\tau _3}}
\end{array}
\right)\,,
\end{equation}
where we kept only the leading order term in the volume expansion for each entry.
Treating both the volume and the blow-up cycle as stable during inflation,
the kinetic term for the fibre modulus simplifies to:
\be
\mc{L}_{\rm kin}=\frac{3}{8\tau_1^2}\partial_\mu \tau_1 \partial^\mu \tau_1\,.
\ee
It then follows that the canonically normalised field is defined as:
\be
\phi\equiv \frac{\sqrt{3}}{2} \ln \tau_1\qquad \text{or} \qquad \tau_1\equiv e^{2\phi/\sqrt{3}}.
\label{eq:phitau11}
\ee
These definitions imply that the shift $\hat \phi$ of the canonically normalised inflaton from its minimum
takes the form:
\be
\hat \tau_1=\langle\tau_1\rangle\left(e^{\frac{2}{\sqrt 3}
\hat \phi}-1\right)\,,
\label{eq:phitau12}
\ee
and so the full inflationary potential (\ref{eq:Vinf}) can be rewritten as:
\be
V_{\rm inf}= \frac{F_{\rm poly}}{\langle\vo\rangle^{3+p}} \left[1-\,e^{-c\left(e^{\frac{2}{\sqrt 3}
\hat \phi}-1\right)}\left(1+c\left(\,e^{\frac{2 \hat \phi}{\sqrt 3}}-1\right)\right)\right]\,,
\label{eq:VCanNorm}
\ee
where $c=2\pi \langle\tau_1\rangle$. This potential is sketched in Figure \ref{fig:V}.
The relation (\ref{eq:phitau12}) suggests the following field redefinition:
\be
\hat\psi\equiv \frac{\sqrt 3}{2}\left(e^{\frac{2}{\sqrt 3}\hat \phi}-1\right)\qquad\Leftrightarrow\qquad
\hat\tau_1=\frac{2}{\sqrt 3}\langle\tau_1\rangle\,\hat\psi\,,
\label{Redef}
\ee
which allows us to write (\ref{eq:VCanNorm}) in a very compact form as:
\be
V_{\rm inf}\simeq \frac{F_{\rm poly}}{\langle\vo\rangle^{3+p}}
\left(1-\kappa\,\hat\psi\,e^{-\kappa\,\hat\psi}\right)\,,
\label{Vinf}
\ee
where $\kappa= \frac{2 p}{\sqrt 3}\ln\vo$ and we have approximated $c = p\ln\vo+1-p\ln f_2\simeq p\ln\vo$.
We stress that we shall obtain a model of small field inflation where
the inflaton travels a sub-Planckian distance in field space during inflation:
$\Delta \hat\phi\simeq 0.5 M_p$. Therefore we can Taylor expand the
exponent in (\ref{Redef}) finding that the field $\hat\psi$ gives
the leading order approximation of the canonically normalised inflaton $\hat\phi$
since:
\be
\hat\psi\equiv \frac{\sqrt 3}{2}\left(e^{\frac{2}{\sqrt 3}\hat \phi}-1\right)
\simeq \hat \phi + \frac{1}{\sqrt 3}\,\hat \phi^2 + ...\,.
\label{apppf}
\ee
Thus the compact expression (\ref{Vinf}) gives a rather accurate qualitative description of
the inflationary dynamics when we just identify $\hat\psi$ with $\hat\phi$.

The aim of this work is to investigate whether
the dynamics of the scalar field $\hat \phi$ rolling down the potential (\ref{eq:VCanNorm}) is suitable for
generating a prolonged period of inflation.
With this in mind we compute the slow-roll parameters for the canonically normalised fibre modulus:
\be
\epsilon=\frac{1}{2 V_{\rm inf}^2}\left(\frac{\partial V_{\rm inf}}{\partial \phi}\right)^2
\qquad \text{and} \qquad \eta=\frac {1}{V_{\rm inf}}\frac{\partial^2 V_{\rm inf}}{\partial \phi ^2}\,.
\ee
The slow-roll parameters turn out to be (for simplicity we express them in terms of $\hat \psi$):
\be
\epsilon=\frac{2 c^2}{3} \frac{\left(\frac{2 c}{\sqrt{3}} \hat\psi\right)^2 \left(\frac{2}{\sqrt{3}} \hat\psi
   +1\right)^2}{\left(e^{\frac{2 c}{\sqrt 3}\hat\psi }-1-\frac{2 c}{\sqrt{3}} \hat\psi \right)^2}
   \simeq \kappa^4 \,e^{-2 \kappa \hat\psi}\,,
\label{eps}
\ee
and:
\be
\eta=-\frac{4c}{3} \frac{\left(\frac{2}{\sqrt 3}\hat\psi +1\right) \left[\left(\frac{2 c}{\sqrt 3} \hat\psi\right) ^2
+(c-2)\frac{2 c }{\sqrt 3}\hat\psi-c\right]}
{e^{\frac{2 c}{\sqrt 3}\hat\psi}-1-\frac{2 c }{\sqrt 3}\hat\psi }\simeq
- \kappa^3 \, e^{-\kappa\hat\psi}\,.
\label{eta}
\ee
Both $\epsilon$ and $\eta$ are naturally exponentially small in the region $\kappa\hat\psi\gg 1$.
Due to the large parameter $\kappa= \frac{2 p}{\sqrt 3}\ln\vo\gg 1$, the slow-roll conditions are
satisfied for small shifts $\hat\psi< 1$. This implies that we are dealing with a model
of small field inflation and justifies our leading order approximation $\hat\psi\simeq \hat\phi$.
Moreover our model is characterised by the interesting relation:
\be
\epsilon\simeq \left(\frac{\eta}{\kappa}\right)^2\,,
\ee
which implies the following hierarchy throughout all the inflationary region:
\be
\epsilon\ll|\eta|\ll 1\,.
\label{eq:SlowRoll}
\ee
Requiring that $|\eta|$ is at most a few percents leads to values of $\epsilon$ within the interval
$10^{-7}<\epsilon<10^{-5}$. Notice that a negative $\eta$ is a generic feature of our
potential showing that the inflating region is tachyonic and therefore unstable.
Hence the field $\hat \phi$ drives inflation as it slowly rolls towards its minimum.

\subsection{Inflationary observables}

In order to solve the basic problems of standard Big-Bang cosmology, the inflationary trajectory
must also give rise to a sufficient number of e-foldings:
\be
N_e\equiv \int_{\hat\phi_{\rm end}}^{\hat\phi_*} \frac{1}{\sqrt{2\epsilon}} d \hat\phi
\simeq \int_{\hat\phi_{\rm end}}^{\hat\phi_*} \frac{\kappa}{|\eta|} d \hat\phi \simeq
\frac{1}{\kappa^2}\int_{\hat\phi_{\rm end}}^{\hat\phi_*} e^{\kappa\hat\phi} d \hat\phi
\simeq \frac{1}{\kappa^3}\left(e^{\kappa\hat\phi_*}-e^{\kappa\hat\phi_{\rm end}}\right)\,.
\label{Ne}
\ee
In our numerical analysis we shall consider the end of inflation
as taking place at the point $\hat\phi_{\rm end}$ where $\epsilon\simeq 1$.
We stress that there is nothing special in this choice since our final results
for the cosmological observables are not sensitive to the exact point where inflation ends.
From the expression (\ref{eps}) we can then see that $\hat\phi_{\rm end}$ is, in practice,
just a function of $\kappa$ since it is rather insensitive to the other underlying parameters.
In turn, from (\ref{Ne}), we see that the number of e-foldings $N_e$ is a function of $\kappa$
and the point of horizon exit $\hat\phi_*$.

However, the exact number of e-foldings depends on the inflationary scale
$M_{\rm inf}=V_{\rm inf}^{1/4}(\hat\phi_{\rm end})$
and the reheating temperature $T_{\rm rh}$. In fact,
under the general assumption that inflation is followed first by a
matter-dominated reheating epoch, and then by a radiation-dominated
epoch with initial temperature $T_{\rm rh}$,
we have \cite{KolbandTurner}:
\be
 N_e\simeq 62+\ln\left(\frac{M_{\rm inf}}{10^{16}\,
 {\rm GeV}}\right)-\frac 13
 \ln\left(\frac{M_{\rm inf}}{T_{\rm rh}}\right)\,.
 \label{cosmology}
\ee
As can be seen from (\ref{eq:VCanNorm}), $M_{\rm inf}$ depends in general on $\vo$,
$p$ and other parameters via the combination $F_{\rm poly}$ of Eq. (\ref{eq:Vinf}).
For fixed $F_{\rm poly}$, we can consider $M_{\rm inf}$ as depending on just $\kappa$
given that the inflationary scale is rather insensitive to the actual value of $p$
which we shall always set $p\simeq \mc{O}(1)$.
As we shall argue later on, $T_{\rm rh}$ is also a function of $\kappa$.
Hence by equating (\ref{Ne}) with (\ref{cosmology}), we obtain an equation in
two unknowns: $\hat\phi_*$ and $\kappa$ (or $\vo$).

We can find a unique solution for each value of $F_{\rm poly}$ by noticing that the requirement of
generating the correct amount of density perturbations
gives a second equation in $\hat\phi_*$ and $\kappa$.
In fact, the COBE normalisation
can be written in terms of the inflaton potential as \footnote{We included a prefactor of $g_s/8\pi$
for the correct normalisation of the potential in Einstein frame \cite{Burgess:2010bz}.}:
\be
A_{\rm COBE}\equiv \frac{g_s}{8\pi}\left.\left(\frac{V_{\rm inf}^{3/2}}{V_{\rm inf}'}\right)^2\right|_{\hat\phi=\hat\phi_*}
\simeq 2.7 \cdot 10^{-7}\,.
\label{eq:COBE}
\ee
We solved numerically these two equations for $p\simeq \mc{O}(1)$ and different values of $F_{\rm poly}$.
We found solutions with an internal volume large enough to trust the effective field theory,
$\vo\sim 10^3$, corresponding to $\kappa\simeq 8$, for $F_{\rm poly}\sim \mc{O}(10)$ which gives also
$\hat\phi_*\simeq 0.9$. 
In turn, these results lead to $\hat\phi_{\rm end}\simeq 0.35$,
$N_e\simeq 54$ and $M_{\rm inf}\simeq M_p/\vo\simeq 10^{15}$ GeV. Notice that these numerical results confirm
our initial statement that we are dealing with a model of small field inflation since
$\Delta\hat\phi=\hat\phi_*-\hat\phi_{\rm end}\simeq 0.45$ in Planck units.

Since our model has a preference for moderately small volumes
it seems hard to simultaneously accommodate GUT-scale inflation and TeV-scale SUSY
which would require $\vo\sim 10^{15}$. This is the infamous tension between the scale of inflation
and the scale of SUSY breaking \cite{Kallosh:2004yh} which afflicts most of the models of string inflation.
Given that in this work we are interested only in inflation,
we shall not attempt to address this issue (see \cite{VolInfl,He:2010uk} for possible solutions).

Let us now turn to the study of the observational footprints of our model.
The spectral index and the tensor-to-scalar ratio are defined by:
\be
n_s=1+2\eta_*-6\epsilon_* \qquad \text{and}\qquad r=16 \epsilon_*\,,
\ee
and they take the values:
\be
n_s\simeq 0.96\qquad \text{and}\qquad r\simeq 10^{-5}\,.
\ee
Notice that since $\eta<0$ this inflationary model necessarily leads to the observationally preferred
value $n_s<1$. The smallness of $\epsilon$ implies the absence of observable primordial tensor modes
in agreement with the fact that the inflaton range during inflation is sub-Planckian \cite{Lyth}.
In section \ref{sec:stringloops}, after discussing the effect of loop corrections on
the inflationary potential, we shall present three different numerical fits for the underlying parameters
which lead to these successful predictions in agreement with current data.

\subsection{Reheating}

The study of reheating for models of closed string inflation has already been performed in \cite{Kofman,Cicoli:2010ha}.
After the end of inflation, the inflaton behaves as a classical condensate
which oscillates coherently around its minimum. Due to the steepness of the
potential, these oscillations end very rapidly because of the non-perturbative
production of inflaton quanta \cite{Kofman}. Subsequently reheating takes place via the perturbative decay
of inflaton particles into visible sector degrees of freedom localised on D7-branes wrapping
an additional blow-up mode \cite{Cicoli:2010ha}.

The gravitational coupling of the inflaton to all the other particles in the model
can be read off from the moduli dependence of
the kinetic and mass terms of open string modes by expanding the moduli around their VEVs
and then expressing them in terms of the canonically normalised fields \cite{Couplings}.
Following this procedure, the inflaton coupling to visible
sector gauge bosons scales with the overall volume as \cite{ADDstrings}:
\be
g \simeq \frac{1}{M_p \vo^p}\,.
\ee
Notice that this coupling is much weaker than gravitational, reflecting
the fact that the direction $\tau_1$ is not lifted by the leading order stabilising dynamics.
Thus the width of the inflaton decay into visible sector degrees of freedom behaves as:
\be
\Gamma \simeq g^2 m_{\tau_1}^3 \simeq \frac{M_p}{\vo^{(9+ 7 p)/2}}\,,
\ee
and so the corresponding reheating temperature is given by:
\be
T_{\rm rh}\simeq \sqrt{\Gamma M_p} \simeq \frac{M_p}{\vo^{(9+ 7 p)/4}}\,.
\label{Trh}
\ee
For $p\simeq \mc{O}(1)$ and $\vo \sim 10^3$, this expression gives
$T_{\rm rh}\simeq 10^6$ GeV which is much higher than the
Big-Bang Nucleosynthesis temperature $T_{\rm BBN}\simeq 1$ MeV,
and moreover it allows electro-weak baryogenesis.
Notice that the inflaton dumps all its energy into visible degrees of freedom
since its decay to hidden sector particles is kinematically forbidden.
In fact, the condensing field theory on $\tau_3$ develops a mass gap and
all the particles acquire a mass of the order the scale of strong dynamics
$\Lambda\sim M_p /\vo^{5/6}\gg m_{\tau_1}$ \cite{Cicoli:2010ha}.

\subsection{Loop corrections}
\label{sec:stringloops}

In the discussion of the structure of the scalar potential for the fibre modulus we have
so far neglected the effect of $g_s$ corrections to the K\"ahler potential.
In this section we shall therefore study their behaviour showing that they are naturally
subleading with respect to the poly-instanton corrections both around the minimum and,
more importantly, in all the inflationary region. Thus our model features a nice solution
of the $\eta$-problem.

First of all, a key observation is that the presence of open string loop corrections
would definitely dominate over the tiny poly-instanton effects, and generate a
potential for $\tau_1$ which is also able to give rise to slow-roll inflation
due to the `extended no-scale structure' as studied in \cite{FI}. We shall
however forbid the presence of these $g_s$ corrections to $K$ by
not wrapping any D7-brane either on $\tau_1$ or on any four-cycle intersecting
the K3 or $T^4$ fibre (i.e. $\tau_2$ in our case). In this way there is
no open string localised on $\tau_1$, and so we do not expect any $\tau_1$-dependent
open string loop correction to $K$.

However, closed string loops might still introduce a dependence on the fibre modulus
since they correspond to loops of bulk Kaluza-Klein modes which cannot be avoided
by construction. Hence it is crucial to study the behaviour of these $g_s$ effects
and their relative strength with respect to poly-instanton corrections.
Following \cite{Cicoli:2007xp}, the dependence of these corrections on the K\"ahler moduli
can be estimated by using the one-loop Coleman-Weinberg potential \cite{ColeWbg}:
\be
\delta V_{(g_s)} \simeq \Lambda^4\,{\rm STr}(M^0) + \Lambda^2\, {\rm STr} (M^2)\,,
\label{CWpot}
\ee
where the first contribution vanishes since supersymmetry implies ${\rm STr}(M^0) = 0$.
The cut-off $\Lambda$ can be taken as the scale at which the 4D effective field theory description ceases to be valid. Given the hierarchy of scales described in \cite{ADDstrings} this should be $M_{KK}^{6D}$, the scale above which the theory becomes 6D:
\be
\Lambda \equiv M_{KK}^{6D}\simeq \frac{M_s}{t_1^{1/2}}\simeq \frac{ \sqrt{\tau_1}}{\mc{V}}M_p.
\ee
The supertrace can instead be approximated as ${\rm STr}(M^2)\simeq m_{3/2}^2\simeq \left(W_0 M_p/\vo\right)^2$,
and so the expression (\ref{CWpot}) takes the final form:
\be
\delta V_{(g_s)} \simeq \left(g_s C_{\rm loop}\right)^2 W_0^2 \frac{\tau_1}{\vo^4}\,,
\label{eq:Vloop}
\ee
where $C_{\rm loop}$ is an unknown function of the complex structure moduli
which we had to introduce by hand since our argument does not allow us
to constrain the dependence of the one-loop potential on the $U$-moduli \footnote{The factor of $g_s$ can be
derived by demanding that the one-loop corrected $K$ scales as $g_s^2$ in string frame.}.
However, this is not a problem since the complex structure moduli are flux-stabilised at tree-level,
and so we can consider $C_{\rm loop}$ just as a constant parameter.
This parameter is squared due to the extended no-scale structure \cite{Cicoli:2007xp}
which implies the vanishing of the leading contribution proportional to $C_{\rm loop}$
(this corresponds to the vanishing of the $\mc{O}(\Lambda^4)$ term in the Coleman-Weinberg potential).

Therefore the loop-corrected potential for the canonically normalised inflaton becomes:
\be
V= \frac{F_{\rm poly}}{\langle\vo\rangle^{3+p}} \left[1-\,e^{-c\left(e^{\frac{2}{\sqrt 3}
\hat \phi}-1\right)}\left(1+c\left(\,e^{\frac{2 \hat \phi}{\sqrt 3}}-1\right)\right)\right]
+\frac{F_{\rm loop}}{\langle\vo\rangle^4}\,e^{\frac{2}{\sqrt 3}\hat \phi}\,,
\ee
where we recall that $c=2 \pi \langle \tau_1\rangle$, and we have defined:
\be
F_{\rm loop}= \frac{c}{2\pi}\left(g_s C_{\rm loop} W_0\right)^2\,.
\label{Floop}
\ee

\begin{figure}[h]
\begin{center}
\includegraphics[width=7cm]{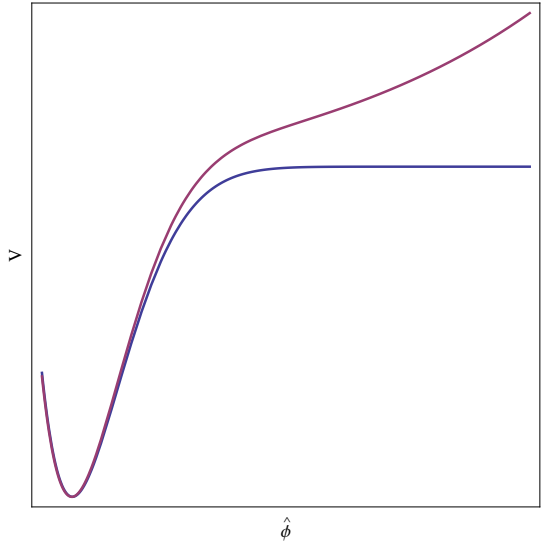}
\caption{Sketch of the inflationary potential for an illustrative choice of the underlying parameters.
In blue: the potential considering
only the contribution from poly-instantons;
in magenta: the potential including loop corrections.}
\label{fig:V}
\end{center}
\end{figure}

As can be seen from Figure \ref{fig:V}, the effect of the one-loop correction
is to lift the inflationary region and to generate
an inflection point at $\hat\phi=\hat\phi_{\rm ip}$ where
$\partial ^2 V/\partial \hat\phi^2=0$.
The position of the inflection point as a function of the ratio
 \be R \equiv F_{\rm loop}/F_{\rm poly}\ee
can be approximated by \footnote{We again use $\hat\psi$ instead of $\hat\phi$ for simplicity.
Recall that at leading order the two
quantities coincide (see Eq. (\ref{apppf})).}:
\be
e^{\kappa\hat\psi_{\rm ip}}\simeq \frac{\kappa^3\langle\vo\rangle^{1-p}}{R}\,,
\label{eq:phiip}
\ee
showing correctly that $\hat\phi_{\rm ip}$ gets larger for smaller values of $R$, i.e. when
the loops get weaker than the poly-instanton effects.
The position of the inflection point is also dependent on the value of the
overall volume via the parameter $p$, showing that the loop corrections
get volume suppressed for $p<1$. We shall however focus on models
where $p\lesssim 1$ since values of $p$ much smaller than unity would both
render the inflaton heavier and bring $\langle\tau_1\rangle$ in a regime
where we no longer trust the effective field theory.

For $\hat\phi>\hat\phi_{\rm ip}$
the slow-roll parameter $\eta$ becomes positive while $\epsilon$ stays negligibly small.
Hence if inflation started at $\hat\phi_*>\hat\phi_{\rm ip}$,
we would obtain a spectral index $n_s>1$ which is incompatible with observations.
We need therefore to focus on cases
where the ratio $R$ is small enough to have the inflection point
lying outside the inflationary region, i.e. $\hat\phi_{\rm ip}>\hat\phi_*$.
In this way, we can realise inflation with a potential generated
by the poly-instantons in a region in which loop effects are negligible.

\begin{figure}[h!]
	\centering
	\begin{minipage}[b]{0.31\linewidth}
	\centering
	\includegraphics[width=\textwidth]{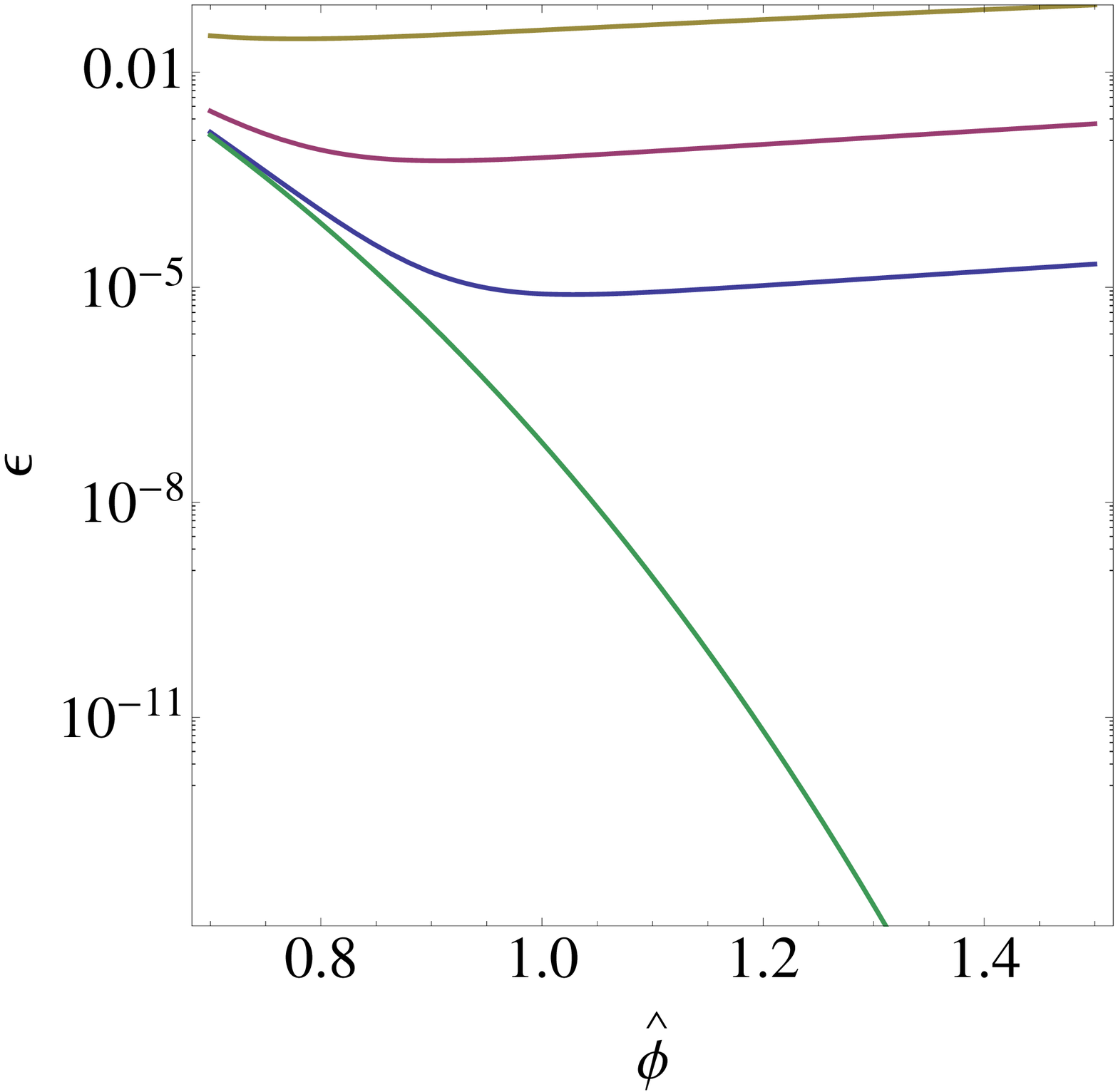}
    \end{minipage}
	\hspace{0.1cm}
	\begin{minipage}[b]{0.31\linewidth}
	\centering
	\includegraphics[width=\textwidth]{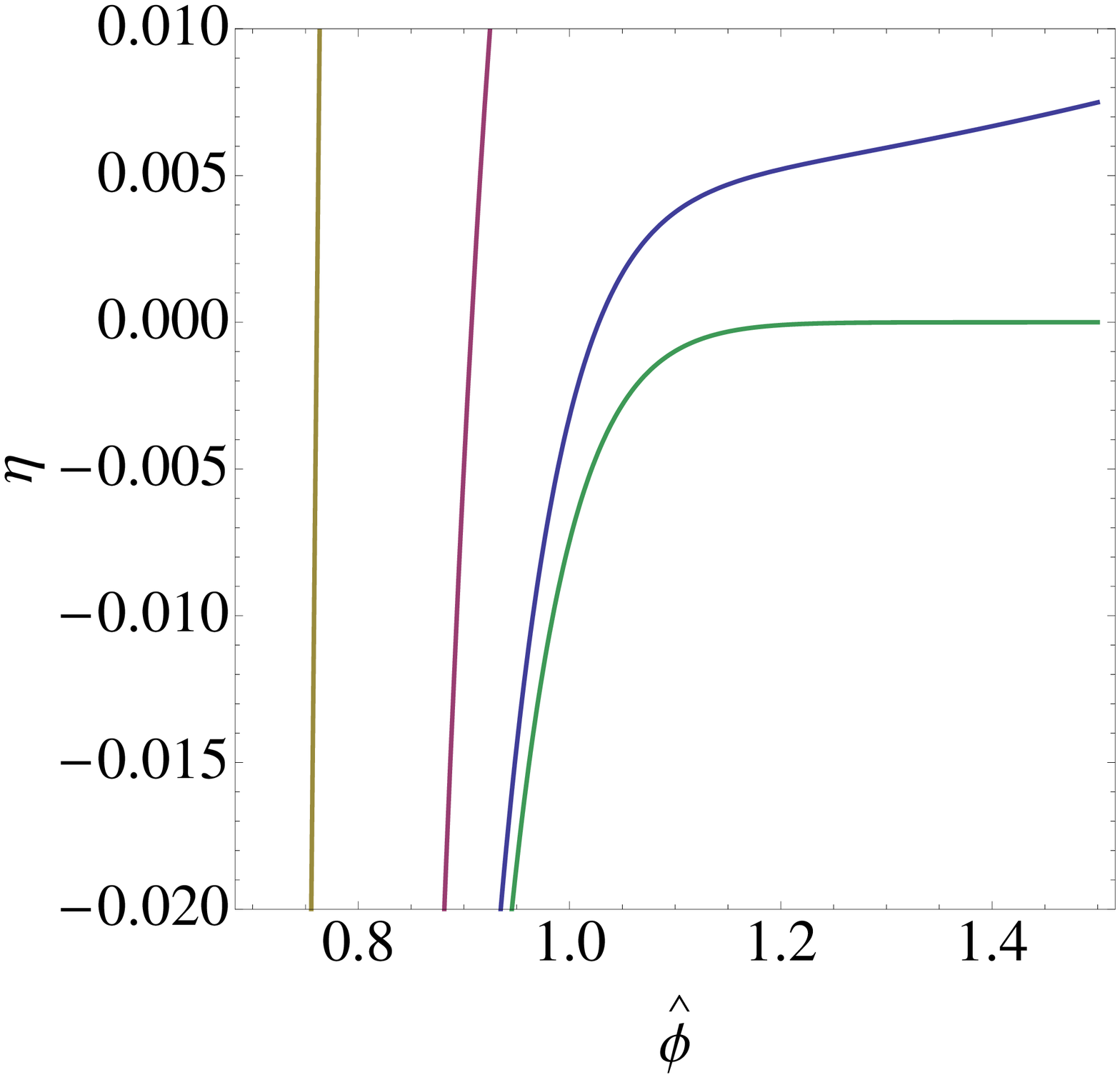}
	\end{minipage}
	\hspace{0.1cm}
	\begin{minipage}[b]{0.31\linewidth}
	\centering
	\includegraphics[width=\textwidth]{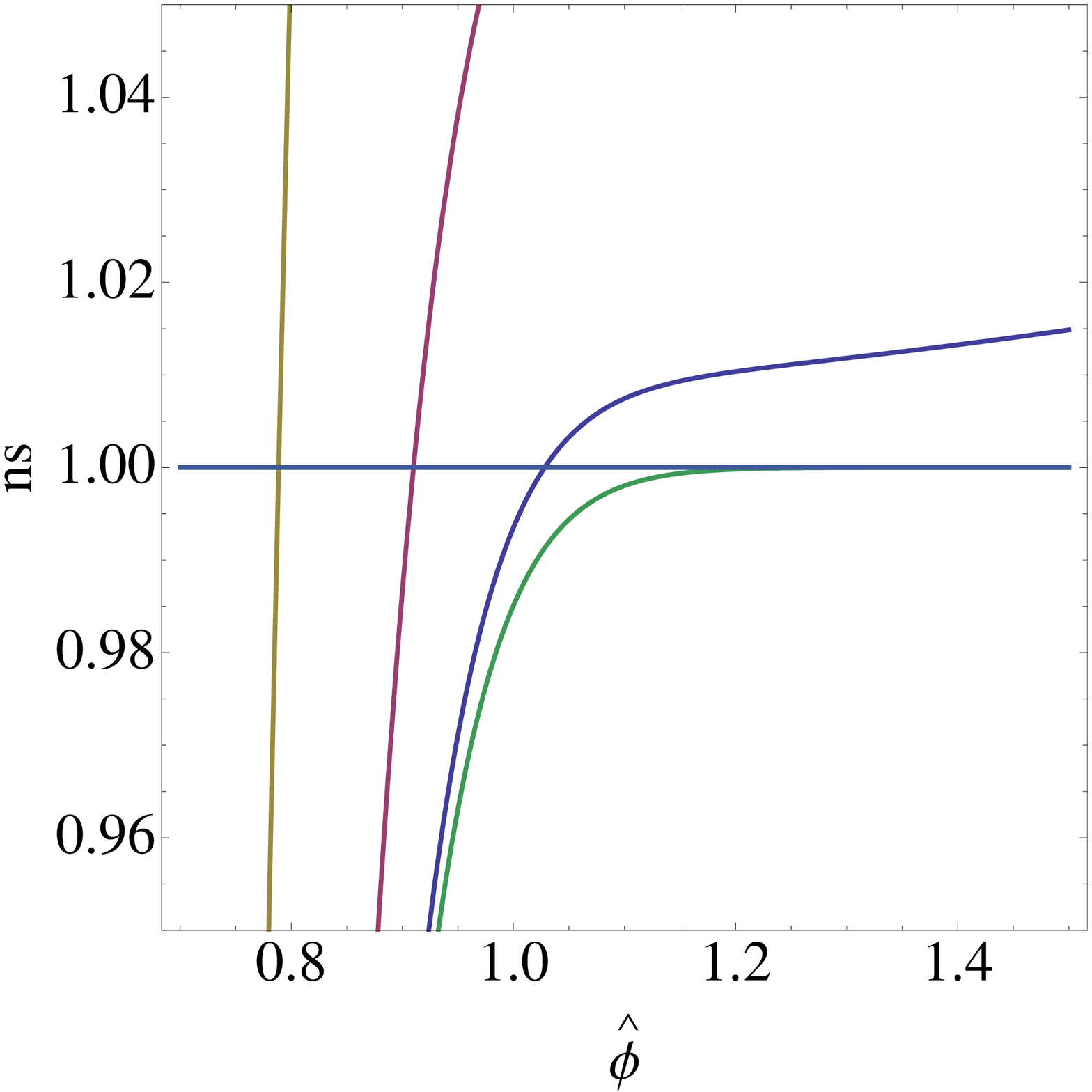}
	\end{minipage}
	\caption{Slow-roll parameters and spectral index for different indicative values of $R$ setting $p=1$:
    blue $R=10^{-3}$, magenta $R=10^{-2}$, yellow $R=10^{-1}$, green $R=0$.}
	\label{fig:nsLoops}
\end{figure}

In Figure \ref{fig:nsLoops} we plot the slow-roll parameters
and the spectral index for different indicative values of $R$.
Given that the numerical study of the potential generated
by poly-instanton effects
gives $\hat\phi_*\simeq 0.9$ for $N_e\simeq 54$,
Figure \ref{fig:nsLoops} reveals that
the effect of loop corrections is negligible for $R\lesssim 10^{-3}$
(corresponding to $\hat\phi_{\rm ip}\gtrsim\hat\phi_*$).

From the definitions of $F_{\rm poly}$, Eq. (\ref{eq:Vinf}), and $F_{\rm loop}$, Eq. (\ref{Floop}),
it follows that $R$ is naturally very small, being proportional to the small parameter $\left(g_s C_{\rm loop}\right)^2\ll 1$:
\be
R\equiv \frac{F_{\rm loop}}{F_{\rm poly}} = \left(g_s C_{\rm loop}\right)^2\left(\frac{W_0\,c}{8\pi f_2^{p+1}r_1\,e^{-1}}\right)\ll 1\,,
\label{ratio}
\ee
since we expect all the parameters in the second parenthesis to be $\mc{O}(1-10)$.
Thus this analysis shows that the loop corrections to the potential are naturally a subleading effect
that does not affect the inflationary dynamics.\\

We present now three illustrative parameter fits which we found numerically.
All of them satisfy all our phenomenological and theoretical constraints and give rise to a viable inflationary model.
Table \ref{tab:parameters} displays the underlying parameters,
Table \ref{tab:volumes} the compactification properties, and Table \ref{tab:inflation}
the inflationary footprints.

\begin{table}[h]
\begin{center}
\begin{tabular}{c|c|c|c|c|c|c|c|c|c|c|c}
&$W_0$ & $A$ & $B$ &$a$ &$b$ & $C_1$ & $n$ &$\xi$ & $g_s$ & $\alpha$&$\beta$\\
\hline
$\mc{P}_1$&$14$&$11$&$8$&$2\pi/4$&$2\pi/5$&$4$&$-2.308$&$1.5$&$0.15$&$\sqrt{1.5}$&$\sqrt{1.5}$\\
$\mc{P}_2$&$6$&$12$&$6$&$2\pi/4$&$2\pi/5$&$3$&$-1.198$&$1.5$&$0.15$&$\sqrt{1.5}$&$\sqrt{1.5}$\\
$\mc{P}_3$&$5$&$30$&$10$&$2\pi/4$&$2\pi/5$&$9$&$-3.090$&$1.5$&$0.2$&$0.717$&$1$\\
\hline
\end{tabular}
\end{center}
\caption{Sampling of points in parameter space.}
\label{tab:parameters}
\end{table}

\begin{table}[h]
\begin{center}
\begin{tabular}{c|c|c|c|c|c|c|c}
 &$\langle \tau_3 \rangle$ & $\langle \mc{V}\rangle$ & $\langle\tau_1\rangle$ &$p$ &$Z$ &$r_1$ &$R/C_{loop}^2$\\
\hline
$\mc{P}_1$&4.57 &$1480$&$1.07$&$1$&$5.95$&$0.58$&$0.017$\\
$\mc{P}_2$&4.68 &$1300$&$1.00$&$0.9$&$3.21$&$0.60$&$0.013$\\
$\mc{P}_3$&5.70 &$1419$&$1.07$&$0.8$&$4.69$&$3.42$&$0.035$\\
\hline
\end{tabular}
\end{center}
\caption{Minima for the K\"ahler moduli generated for the microscopic parameters of Table \ref{tab:parameters}.}
\label{tab:volumes}
\end{table}

\begin{table}[h]
\begin{center}
\begin{tabular}{c|c|c|c|c|c}
& $N_e$ & $n_s$ & $r$ &$A_{\rm COBE}$ &$M_{\rm inf} ({\rm GeV})$  \\
\hline
$\mc{P}_1$& $54.2$ & $0.959$ &$9.56 \times 10^{-6}$& $2.63\times10^{-7}$ &$1.81\times10^{15}$  \\
$\mc{P}_2$& $54.2$ & $0.959$ & $9.56 \times 10^{-6}$&$2.26\times 10^{-7}$ &$1.75\times10^{15}$  \\
$\mc{P}_3$& $54.2$ & $0.959$ & $9.56 \times 10^{-6}$&$2.11\times 10^{-7}$ &$1.72\times 10^{15}$  \\
\hline
\end{tabular}
\end{center}
\caption{Inflationary observables for the fits of Table \ref{tab:parameters}.}
\label{tab:inflation}
\end{table}

Comparing the results of Tables \ref{tab:parameters}, \ref{tab:volumes} and
\ref{tab:inflation} we see that it is possible to generate inflation at the right scale
for natural values of the microscopic parameters, with the fibre modulus stabilised in the geometric regime.
Furthermore we see in Table \ref{tab:volumes} that the values of the ratio $R$ turn
out to be always smaller than $10^{-3}$ for $C_{loop}\sim 0.1$, justifying the fact that we neglected string loop effects.

We stress that this was not possible in the model of \cite{KMI} even if the inflationary
potential was still generated by non-perturbative effects,
and so had a shape very similar to (\ref{Vinf}). The reason is that in \cite{KMI} the parameter $\kappa$
scales with the volume as $\kappa\sim \vo^{1/2}$, so it is large for large values of the volume. Instead,  in our case it
 results much smaller since it scales with the volume  as  $\kappa\sim \ln\vo$. This is due to the different topological
origin of the inflaton which in \cite{KMI} was a blow-up mode whereas,  in our case it is
a fibre modulus. The fact that $\kappa$ is smaller in our set-up, with respect to the model of \cite{KMI},
has a two-fold implication:
\begin{itemize}
\item Our potential is steeper than the one of \cite{KMI} resulting in a larger value of
the tensor-to-scalar ratio $r\sim 10^{-5}$ compared to the value $r\sim 10^{-10}$ of \cite{KMI};

\item Looking at the dependence of the slow-roll parameters (\ref{eps}) and (\ref{eta}) on $\kappa$,
we realise that larger values of $\kappa$ require larger values of $\hat\phi$ in order to satisfy the
slow-roll conditions. Hence the inflationary region in our case is much closer to the
minimum than in \cite{KMI}. This is the reason why, contrary to \cite{KMI}, our model is not
affected by the $\eta$-problem since we can avoid to experience regions in field space
that are affected by string loop corrections.
\end{itemize}
In our set-up, our numerical analysis confirms that our potential
gives rise to negligible tensor contributions to the CMB,  and with
a spectral index for scalar perturbations that is compatible with current observations. 
Notice that the fibre inflation model of \cite{FI} has also an inflationary potential with a shape very similar to (\ref{Vinf}).
However in \cite{FI} the parameter $\kappa$ is smaller and volume-independent: its value is  $\kappa=1/\sqrt{3}$. 
This gives a steeper potential with respect to our case,  which is able to give rise to observable tensor modes.

\section{Conclusions}
\label{Concl}

In this paper we derived a new model of inflation driven by a K\"ahler modulus
within the context of type IIB Calabi-Yau flux compactifications.
This class of inflationary models features an interesting solution to the $\eta$-problem
since they are characterised by the presence of several directions which are naturally flat.
In fact, due to the no-scale structure of the K\"ahler potential,
all the $h^{1,1}$ directions in moduli space are flat at tree-level.
Then the leading order effect, that breaks the no-scale structure,
is an $\alpha'$ correction which lifts only the volume mode.
Hence all the other $(h^{1,1}-1)$ directions in moduli space are
flatter than the volume mode which sets the Hubble scale $H$.
This implies that all these scalar fields are natural inflaton candidates
since they turn out to be lighter than $H$.
Moreover, the expansion of the prefactor $e^K$ of the F-term scalar potential
generates higher-dimensional operators which depend only on the volume mode,
and so do not introduce any dangerously large correction to the inflaton mass.

An inflationary potential for these remaining flat directions can be
generated at subleading order either
by non-perturbative corrections to the superpotential, as in K\"ahler moduli inflation \cite{KMI}, or by
perturbative string loop corrections to the K\"ahler potential, as in fibre inflation \cite{FI}.
However, as noticed in \cite{FI},
the model of \cite{KMI} is plagued by the $\eta$-problem since loop corrections dominate over
the non-perturbative potential and spoil inflation.

In this paper we proposed a new way to generate an inflationary potential via
non-perturbative effects which does not lead to an $\eta$-problem.
The new ingredient is the inclusion of poly-instanton corrections to the superpotential \cite{Blumenhagen:2008ji}
for Calabi-Yau manifolds with a K3 or $T^4$ fibre over a $\mbb{P}^1$ base \cite{ADDstrings, Cicoli:2011it}.
The main difference with the model of \cite{KMI} is the topological
nature of the inflaton field. In our case it is a fibre modulus, whereas in
the case of \cite{KMI} it was a blow-up mode. Due to this difference,
our potential is not affected by the $\eta$-problem.
In fact, in our case, the inflationary region
is much closer to the minimum where the `extended no-scale structure' \cite{Cicoli:2007xp}
guarantees that string loop corrections are subdominant
for natural values of the underlying parameters.
Moreover, the different topology of the inflaton field is reflected also in a
steeper potential with respect to K\"ahler moduli inflation.
We therefore obtain a larger value for the tensor-to-scalar ratio $r\sim 10^{-5}$
compared to the value $r\sim 10^{-10}$ of \cite{KMI}, even if it is still beyond
the observational reach.

This small value of $r$ corresponds to a high inflationary scale, $M_{\rm inf}\simeq 10^{15}$ GeV,
and a sub-Planckian motion of the inflaton in field space $\Delta\phi\simeq 0.5 M_p$.
Horizon exit takes place around $N_e\simeq 54$ e-folding before the end of inflation where the spectral index
assumes a value $n_s\simeq 0.96$ in perfect agreement with current observations.
At the end of inflation, the visible sector degrees of freedom get excited by
the gravitational decay of the inflaton field which leads to a reheating temperature of the order
$T_{\rm rh}\simeq 10^6$ GeV.

Finally it is worth emphasising that our analysis of the effect of string loop 
corrections on the inflationary potential relies on a low-energy estimate 
instead of a proper computation of string scattering amplitudes. 
Due to the relatively simple dependence of these loop corrections on the K\"ahler moduli, 
we believe the results we find here are likely to capture their leading behaviour. 
However, it would be very interesting to have more explicit calculations
of $g_s$ corrections to the K\"ahler potential for general Calabi-Yau manifolds.

Apart from  inflation, 
we point out that our set-up possesses two crucial properties that can be applied to build  models for
 quintessence. The first is that our potential
is unaffected by loop corrections, and the second is that
the coupling of the inflaton field to visible sector degrees of freedom is weaker than Planck strength.
In this way one could avoid the stringent bounds coming from fifth-force experiments.
Hence it would be interesting to investigate if our framework
 could also give rise to a successful quintessence model.
 We leave this study for future investigation.

\subsection*{Acknowledgments}

We are grateful to Joe Conlon and Fernando Quevedo for interesting discussions and constructive comments.
FGP is supported by Funda\c{c}\~{a}o para a Ci\^{e}ncia e a Tecnologia (Portugal) through the grant SFRH/BD/35756/2007 and by the Theoretical Physics Department of the University of Oxford. MC would like to thank the DESY Theory Group where part of this work has been developed.
GT is supported by an STFC Advanced Fellowship ST/H005498/1.

\bibliographystyle{JHEP}

\begin{thebibliography}{10}


\bibitem{Guth:1980zm}
  A.~H.~Guth,
  Phys.\ Rev.\  D {\bf 23}, 347 (1981).

\bibitem{Linde:1981mu}
  A.~D.~Linde,
  Phys.\ Lett.\  B {\bf 108}, 389 (1982).

\bibitem{Albrecht:1982wi}
  A.~Albrecht and P.~J.~Steinhardt,
  Phys.\ Rev.\ Lett.\  {\bf 48}, 1220 (1982).

\bibitem{etaproblem}
  E.~J.~Copeland, A.~R.~Liddle, D.~H.~Lyth, E.~D.~Stewart and D.~Wands,
  Phys.\ Rev.\  D {\bf 49} (1994) 6410 [arXiv:astro-ph/9401011].

\bibitem{Reviews}
  R.~Kallosh,
  Lect.\ Notes Phys.\  {\bf 738} (2008) 119
  [arXiv:hep-th/0702059];
  F.~Quevedo,
  Class.\ Quant.\ Grav.\  {\bf 19} (2002) 5721
  [arXiv:hep-th/0210292];
  D.~Baumann and L.~McAllister,
  Ann.\ Rev.\ Nucl.\ Part.\ Sci.\  {\bf 59} (2009) 67
  [arXiv:0901.0265 [hep-th]];
  L.~McAllister and E.~Silverstein,
  Gen.\ Rel.\ Grav.\  {\bf 40} (2008) 565
  [arXiv:0710.2951 [hep-th]];


\bibitem{OurReview}
  M.~Cicoli, F.~Quevedo,
  Class.\ Quant.\ Grav.\  {\bf 28 } (2011)  204001.
  [arXiv:1108.2659 [hep-th]].


\bibitem{kklt}
  S.~Kachru, R.~Kallosh, A.~D.~Linde and S.~P.~Trivedi,
  Phys.\ Rev.\  D {\bf 68} (2003) 046005
  [arXiv:hep-th/0301240].

\bibitem{LVS}
  V.~Balasubramanian, P.~Berglund, J.~P.~Conlon and F.~Quevedo,
  JHEP {\bf 0503} (2005) 007
  [arXiv:hep-th/0502058].

\bibitem{Openclosedinflatons}
P.~Binetruy and M.~K.~Gaillard,
  Phys.\ Rev.\  D {\bf 34}, 3069 (1986);
 T.~Banks, M.~Berkooz, S.~H.~Shenker, G.~W.~Moore and P.~J.~Steinhardt,
  Phys.\ Rev.\  D {\bf 52}, 3548 (1995)
  [arXiv:hep-th/9503114].

\bibitem{DDinflation}
G.~R.~Dvali and S.~H.~H.~Tye,
  Phys.\ Lett.\  B {\bf 450} (1999) 72 [arXiv:hep-ph/9812483].

\bibitem{BBbarInfl}
C.~P.~Burgess \textit{et al},
  JHEP {\bf 0107} (2001) 047 [arXiv:hep-th/0105204];
G.~R.~Dvali, Q.~Shafi and S.~Solganik,
arXiv:hep-th/0105203;
C.~Herdeiro, S.~Hirano and R.~Kallosh,
  JHEP {\bf 0112} (2001) 027 [arXiv:hep-th/0110271];
C.~P. Burgess \textit{et al},
JHEP {\bf 0203} (2002) 052 [arXiv:hep-th/0111025];
S.~Kachru \textit{et al},
JCAP {\bf 0310} (2003) 013 [arXiv:hep-th/0308055];
H.~Firouzjahi and S.~H.~H. Tye,
 Phys.\ Lett.\  B {\bf 584} (2004) 147 [arXiv:hep-th/0312020];
N.~Iizuka and S.~P.~Trivedi,
  Phys.\ Rev.\  D {\bf 70} (2004) 043519 [arXiv:hep-th/0403203];
C.~P. Burgess, J.~M. Cline, H.~Stoica, and F.~Quevedo,
JHEP {\bf 09} (2004) 033 [arXiv:hep-th/0403119].

\bibitem{DBI}
 E.~Silverstein and D.~Tong,
  Phys.\ Rev.\  D {\bf 70}, 103505 (2004)
  [arXiv:hep-th/0310221];
M.~Alishahiha \textit{et al},
  Phys.\ Rev.\  D {\bf 70}, 123505 (2004)
  [arXiv:hep-th/0404084];
K.~Becker \textit{et al},
  Nucl.\ Phys.\  B {\bf 715}, 349 (2005)
  [arXiv:hep-th/0501130].


\bibitem{Racetrack}
J.~J.~Blanco-Pillado \textit{et al},
  JHEP {\bf 0411} (2004) 063
  [arXiv:hep-th/0406230];
J.~J.~Blanco-Pillado \textit{et al},
  JHEP {\bf 0609} (2006) 002
  [arXiv:hep-th/0603129].

\bibitem{KMI}
J.~P.~Conlon and F.~Quevedo,
  JHEP {\bf 0601} (2006) 146
  [arXiv:hep-th/0509012];
J.~R.~Bond, L.~Kofman, S.~Prokushkin and P.~M.~Vaudrevange,
  Phys.\ Rev.\  D {\bf 75} (2007) 123511
  [arXiv:hep-th/0612197].

\bibitem{FI}
  M.~Cicoli, C.~P.~Burgess and F.~Quevedo,
  JCAP {\bf 0903} (2009) 013 [arXiv:0808.0691 [hep-th]].

\bibitem{Nflation}
S.~Dimopoulos, S.~Kachru, J.~McGreevy and J.~G.~Wacker,
  JCAP {\bf 0808}, 003 (2008)
  [arXiv:hep-th/0507205];
T.~W.~Grimm,
  Phys.\ Rev.\  D {\bf 77} (2008) 126007
  [arXiv:0710.3883 [hep-th]].

\bibitem{SW}
E.~Silverstein and A.~Westphal,
  Phys.\ Rev.\  D {\bf 78}, 106003 (2008)
  [arXiv:0803.3085 [hep-th]];
L.~McAllister \textit{et al},
  Phys.\ Rev.\  D {\bf 82} (2010) 046003
  [arXiv:0808.0706 [hep-th]].

\bibitem{nemanja}
N.~Kaloper, A.~Lawrence and L.~Sorbo,
  JCAP {\bf 1103} (2011) 023
  [arXiv:1101.0026 [hep-th]].

\bibitem{VolInfl}
  J.~P.~Conlon, R.~Kallosh, A.~D.~Linde and F.~Quevedo,
  JCAP {\bf 0809} (2008) 011
  [arXiv:0806.0809 [hep-th]].

\bibitem{OtherClosedStringInfl}
A.~Avgoustidis, D.~Cremades and F.~Quevedo,
  Gen.\ Rel.\ Grav.\  {\bf 39} (2007) 1203 [arXiv:hep-th/0606031];
M.~Badziak and M.~Olechowski,
  JCAP {\bf 0807}, 021 (2008)
  [arXiv:0802.1014 [hep-th]];
 H.~X.~Yang and H.~L.~Ma,
  JCAP {\bf 0808} (2008) 024 [arXiv:0804.3653 [hep-th]].

\bibitem{shiftInflation}
  J.~P.~Hsu and R.~Kallosh,
  JHEP {\bf 0404} (2004) 042
  [arXiv:hep-th/0402047].

\bibitem{Cicoli:2007xp}
  M.~Cicoli, J.~P.~Conlon and F.~Quevedo,
  JHEP {\bf 0801} (2008) 052
  [arXiv:0708.1873 [hep-th]].

\bibitem{Kallosh:2004yh}
  R.~Kallosh and A.~D.~Linde,
  JHEP {\bf 0412} (2004) 004
  [arXiv:hep-th/0411011].

\bibitem{Blumenhagen:2008ji}
  R.~Blumenhagen and M.~Schmidt-Sommerfeld,
  JHEP {\bf 0807} (2008) 027
  [arXiv:0803.1562 [hep-th]].

\bibitem{ADDstrings}
  M.~Cicoli, C.~P.~Burgess and F.~Quevedo,
  [arXiv:1105.2107 [hep-th]].

\bibitem{Cicoli:2011it}
  M.~Cicoli, M.~Kreuzer, C.~Mayrhofer,
  [arXiv:1107.0383 [hep-th]].

\bibitem{Burgess:2010bz}
  C.~P.~Burgess, M.~Cicoli, M.~Gomez-Reino, F.~Quevedo, G.~Tasinato and I.~Zavala,
  JHEP {\bf 1008} (2010) 045
  [arXiv:1005.4840 [hep-th]].


\bibitem{Grimm:2004uq}
  T.~W.~Grimm and J.~Louis,
  Nucl.\ Phys.\  B {\bf 699} (2004) 387
  [arXiv:hep-th/0403067].


\bibitem{Gukov:1999ya}
  S.~Gukov, C.~Vafa and E.~Witten,
  Nucl.\ Phys.\  B {\bf 584} (2000) 69
  [Erratum-ibid.\  B {\bf 608} (2001) 477]
  [arXiv:hep-th/9906070].

\bibitem{Giddings:2001yu}
  S.~B.~Giddings, S.~Kachru and J.~Polchinski,
  Phys.\ Rev.\  D {\bf 66} (2002) 106006
  [arXiv:hep-th/0105097].

\bibitem{BBHL}
  K.~Becker, M.~Becker, M.~Haack and J.~Louis,
  JHEP {\bf 0206} (2002) 060
  [arXiv:hep-th/0204254].


\bibitem{Palti:2008mg}
  E.~Palti, G.~Tasinato, J.~Ward,
  JHEP {\bf 0806 } (2008)  084.
  [arXiv:0804.1248 [hep-th]].



\bibitem{08051029}
  M.~Cicoli, J.~P.~Conlon and F.~Quevedo,
  JHEP {\bf 0810} (2008) 105
  [arXiv:0805.1029 [hep-th]].

\bibitem{Schulz}
K.~Oguiso,
``On Algebraic Fiber Space Structures on a Calabi-Yau 3-fold,''
  Int.\ J.\  of Math. {\bf 4} (1993) 439-465;
M.~B.~Schulz,
  ``Calabi-Yau duals of torus orientifolds,''
  JHEP {\bf 0605} (2006) 023
  [arXiv:hep-th/0412270].

\bibitem{CMV}
M.~Cicoli, C.~Mayrhofer, R.~Valandro,
  [arXiv:1110.3333 [hep-th]].


\bibitem{He:2010uk}
  T.~He, S.~Kachru and A.~Westphal,
  JHEP {\bf 1006}, 065 (2010)
  [arXiv:1003.4265 [hep-th]].

\bibitem{KolbandTurner}
  E.~W.~Kolb and M.~S.~.~Turner,
  ``The Early Universe,''
{\it Westview Press,1990}.

\bibitem{Kofman}
  N.~Barnaby, J.~R.~Bond, Z.~Huang, L.~Kofman,
  JCAP {\bf 0912 } (2009)  021.
  [arXiv:0909.0503 [hep-th]].

\bibitem{Cicoli:2010ha}
  M.~Cicoli and A.~Mazumdar,
  JCAP {\bf 1009} (2010) 025
  [arXiv:1005.5076 [hep-th]];
M.~Cicoli and A.~Mazumdar,
  Phys.\ Rev.\  D {\bf 83} (2011) 063527
  [arXiv:1010.0941 [hep-th]].

\bibitem{Couplings}
 J.~P.~Conlon, F.~Quevedo,
  JCAP {\bf 0708 } (2007)  019.
  [arXiv:0705.3460 [hep-ph]];
L.~Anguelova, V.~Calo, M.~Cicoli,
  JCAP {\bf 0910 } (2009)  025.
  [arXiv:0904.0051 [hep-th]].

\bibitem{Lyth}
 D.H. Lyth, Phys. Rev. Lett. {\bf 78} (1997) 1861,
 arXiv:hep-ph/9606387.

\bibitem{ColeWbg}
  S.~R.~Coleman, E.~J.~Weinberg,
  ``Radiative Corrections as the Origin of Spontaneous Symmetry Breaking,''
  Phys.\ Rev.\  {\bf D7 } (1973)  1888-1910.



\end{thebibliography}

\end{document}